# Uniformity of fuel target implosion in Heavy Ion Fusion

S. Kawata[1,2], K. Noguchi[1], T. Suzuki[1], T. Karino[1], D. Barada[1,2],
A. I. Ogoyski[3], Y. Y. Ma[1,2]

[1] Graduate School of Engineering, Utsunomiya University, Yohtoh 7-1-2, Utsunomiya 321-8585, Japan

[2] CORE (Center for Optical Research and Education), Utsunomiya University, Yohtoh 7-1-2, Utsunomiya 321-8585, Japan

[3] Department of Physics, Technical University of Varna, Ulitska, Studentska 1, Varna, Bulgaria

**Corresponding Author**: Shigeo Kawata
**Address**: Graduate School of Engineering, Utsunomiya University, Yohtoh 7-1-2, Utsunomiya 321-8585, Japan
**Tel. & Fax**: +81-28-689-6080
 **e-mail**: kwt@cc.utsunomiya-u.ac.jp

**Abstract**

In inertial confinement fusion the target implosion non-uniformity is introduced by a driver beams' illumination non-uniformity, a fuel target alignment error in a fusion reactor, the target fabrication defect, et al. For a steady operation of a fusion power plant the target implosion should be robust against the implosion non-uniformities. In this paper the requirement for the implosion uniformity is first discussed. The implosion uniformity should be less than a few percent. A study on the fuel hotspot dynamics is also presented and shows that the stagnating plasma fluid provides a significant enhancement of vorticity at the final stage of the fuel stagnation. Then non-uniformity mitigation mechanisms of the heavy ion beam (HIB) illumination are also briefly discussed in heavy ion inertial fusion (HIF). A density valley appears in the energy absorber, and the large-scale density valley also works as a radiation energy confinement layer, which contributes to a radiation energy smoothing. In HIF a wobbling heavy ion beam illumination was also introduced to realize a uniform implosion. In the wobbling HIBs illumination, the illumination non-uniformity oscillates in time and space on a HIF target. The oscillating-HIB energy deposition may contribute to the reduction of the HIBs' illumination non-uniformity by its smoothing effect on the HIB illumination non-uniformity and also by a growth mitigation effect on the Rayleigh-Taylor instability.

# 1. INTRODUCTION

In inertial confinement fusion, the fuel target implosion non-uniformity leads a degradation of fusion energy output. The implosion uniformity requirement is stringent. Therefore, it is essentially important to improve the fuel target implosion uniformity, and the implosion non-uniformity would be induced by the driver beam illumination non-uniformity (Bodner, 1981, Lindl, 1995, Miyazawa et al., 2005, Kawata et al., 2009). The target implosion non-uniformity allowed is less than a few percent in inertial fusion target implosion (Emery et al., 1982, Kawata et al., 1984). In heavy ion inertial fusion (HIF) the heavy ion beam (HIB) has preferable features, and the HIB axis is precisely controlled with a high frequency (Arnold et al., 1987, Piritz et al., 2003a, Piritz et al., 2003b, Basko et al., 2004, Logan et al., 2008, Kawata, 2012). The energy efficiency of the HIB generation is high, that would be about 30~40%. The HIBs illumination non-uniformity would be mitigated by a radiation smoothing of the HIB deposition energy and by the wobbling beam motion, that is, the HIB axis oscillation or rotation (Arnold et al., 1982, Piritz et al., 2003a, Piritz et al., 2003b, Basko et al., 2004, Logan et al., 2008, Qin et al., 2010, Kawata, 2012)]. In general, the target implosion non-uniformity is introduced by a driver beams' illumination non-uniformity, an imperfect target sphericity, a non-uniform target density, a target alignment error in a fusion reactor, et al. The target implosion should be robust against the implosion non-uniformities.

The fuel implosion non-uniformity is an important central concern in the HIF imploding fuel plasma. The Rayleigh-Taylor (R-T) instabilities take place at the stages of shock acceleration, steady acceleration and stagnation of the implosion due to additional seeding of R-T instabilities by the imperfections of the HIB energy drive and the target.

In this paper an implosion dynamics is first investigated in heavy ion inertial fusion. Heavy ions deposit their energy inside the target energy absorber, and the energy deposition layer is rather thick, for example, about several hundreds μm or more, depending on the ion particle energy. The requirement for the implosion uniformity is next discussed. The target implosion uniformity should be less than a few percent. Especially the hotspot dynamics is also focused in order to investigate the origin of the large amplitude of the fuel target implosion non-uniformity at the stagnation phase: at the initial stage, a few percent of the implosion non-uniformity would be introduced by various sources of the non-uniformity, including the HIB illumination non-uniformity. During the implosion, the non-uniformity amplitude grows significantly so that the fusion energy output is degraded. At the stagnation phase the fuel mixing is significantly enhanced. In this paper a hot-spot dynamics analysis is introduced at the final stage of the stagnation phase or at just before the fuel ignition, and presents that the fuel fluid vorticity is significantly enhanced during the final stagnation phase, in which the fuel radius shrinks much. The fuel shrinkage at the stagnation phase defines the requirement of the initially introduced non-uniformity. Because the non-uniformity amplitude enhancement, led by the fuel vorticity enhancement, is significant at the final stagnation phase, the initial requirement for the implosion uniformity is stringent.

Then the non-uniformity mitigation mechanisms of the heavy ion beam illumination are discussed in HIF. A density valley appears at the energy absorber, and the density gradient scale length is also thick: about several hundred μm or so. The large density-scale length is unique in HIF and contributes to a reduction of the R-T instability growth rate. In addition, the large-scale density valley also works as a radiation energy confinement layer, which also contributes to a radiation energy smoothing. In HIF a wobbling heavy ion beam illumination was also proposed to realize a uniform implosion. The wobbling HIB axis oscillation is precisely controlled. The oscillating frequency may be several 100MHz~1GHz (Arnold et al., 1987, Piritz et al., 2003a, Piritz et al., 2003b, Basko et al., 2004, Logan et al., 2008, Qin et al., 2010, Kawata, 2012). In the wobbling HIBs illumination, the illumination nonuniformity oscillates in time and space on a HIF target. The oscillating-HIB energy deposition may contribute to the reduction of the HIBs' illumination nonuniformity by its smoothing effect on the HIB illumination nonuniformity and also by a growth mitigation effect on the R-T instability. If the perturbation phase is known, the instability growth can be controlled by a superposition of perturbations; the well-known mechanism is a feedback control to compensate the displacement of physical quantity. If the perturbation is induced by, for example, the HIB axis wobbling, the perturbation phase could be controlled and the instability growth is mitigated by the superposition of the growing perturbations.

## 2. TARGET IMPLOSION DYNAMICS IN HEAVY ION FUSION

A target energy gain required for an energy production is evaluated by a reactor-energy balance in inertial fusion. A driver pulse delivers an energy $E_d$ to a target, which releases fusion energy $E_{fusion}$. The energy gain is $G = E_{fusion}/E_d$. The fusion energy is first converted into electricity by a standard thermal cycle with an efficiency

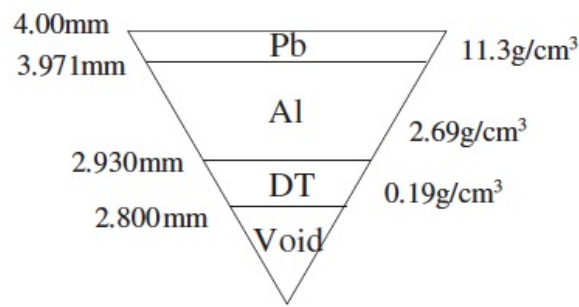

Fig. 1 An example fuel target structure in heavy ion inertial fusion.

of $\eta_{th}$. A fraction $f$ of the electric power is circulated to a driver system, which converts it into HIB energy with an efficiency of $\eta_d$. The energy balance for this cycle is written by $f \eta_{th} \eta_d G = 1$. Taking $\eta_{th} = 40\%$ and requiring that the circulated-energy fraction of electrical energy is less than 1/4, we find the condition $G\eta_d > 10$. For a driver efficiency in the range of $\eta_d = 10 \sim 33\%$, the condition $G = 30 \sim 100$ is required for power production. Therefore, the preferable pellet gain required is about

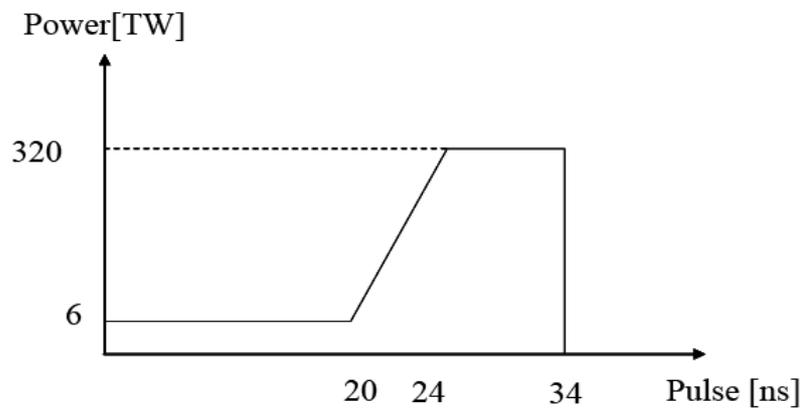

Fig. 2 An input heavy ion beam pulse. The HIB pulse consists of the low power part (foot pulse) and the high power one (main pulse).

$30 \sim 100$ in HIF.

In order to study the fuel target implosion in HIF, we present a target hydrodynamics by using a hydrodynamics code coupling with the HIB illumination code (Someya et al., 2006a, Someya et al., 2006b). We employ a 32-HIBs illumination system (Skupsky et al., 1983). In this paper, two-dimensional simulations are performed. Figure 1 shows the fuel target. The Pb, Al and DT layer thicknesses are 0.03 mm, 0.40 mm, 0.10 mm, and the mass densities 11.3 g/cm$^3$, 2.69 g/cm$^3$ and 0.19 g/cm$^3$, respectively. The HIB input pulse is shown in Fig. 2. In this specific case, the total HIB energy is 4.0 MJ.

Figure 3 presents a mean density and a mean radiation temperature averaged over the θ direction at 36.2 nsec. The averaged HIB illumination non-uniformity is 2.3 % in this case. The 32 Pb ion beams impinge the pellet surface (Skupsky et al., 1983).

The HIB deposition energy distribution produces an ablation region at the Al energy absorber layer, and then about one-third of Al pusher mass pushes the DT fuel. In Fig. 3 the low density region appears at the ablation front. The density gradient scale length of the ablation surface is relatively large in HIF target implosion, that is, about several hundred μm ~ 500μm or so.

When the density gradient scale length $L$ is large, the growth rate ($\gamma$) reduction effect on the R-T instability would be expected (Bodner, 1974, Takabe et al., 1985, Abarzhi, 2010): $\gamma = \sqrt{gk/(1 + kL)}$. Here $g$ is the implosion acceleration, and $k$ the wave number. In HIF, typically $L$ is about several hundred μm ~ 500 μm, and the ablation effect is minor. Therefore, the short wavelength ($2\pi/k$) modes of the perturbation would be suppressed or mitigated by the density gradient effect in HIF. So in HIF typically the large scale perturbation modes, which have the wavelength of several hundred μm ~ 500μm, are serious to keep the fuel target implosion spherically symmetric.

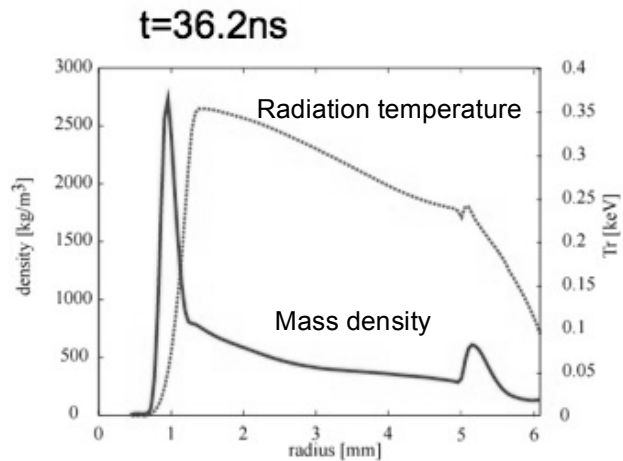

Fig. 3 A mean density and a mean radiation temperature averaged over the θ direction at 36.2 nsec.

## 3. REQUIREMENT OF IMPLOSION UNIFORMITY

In this section, the nonuniform implosion effect is discussed on the target implosion in inertial fusion. It is confirmed that the nonuniformity of the implosion acceleration is required to be less than a few percent in reactor size fuel targets (Emery et al., 1982, Kawata et al., 1984). Therefore, the driver energy deposition should be rather uniform to fulfill the uniform implosion requirement, and smoothing mechanisms are also expected to reduce the implosion non-uniformity. The R-T instabilities must take place at the ablation front and at the stagnation phase in inertial fusion target implosion. For the R-T instability at the stagnation phase, the initial perturbation amplitude and phase are defined by the non-uniform implosion process before the stagnation. The uniform implosion at the acceleration phase is essentially important. The $\rho R$ product of the fuel mass density $\rho$ and the fuel radius $R$ at the stagnation is directly relating to the fusion energy output. When a small implosion non-uniformity is imposed, $\rho R$ is degraded from the perfect uniform $(\rho R)_0$. $\rho R$ is proportional to $1/R^2$. Therefore, $(\rho R)/(\rho R)_0 = \{(R + \delta R)/R\}^{-2} = (1 + \delta R/R)^{-2}$. On the other hand, the nonuniformity $\delta\alpha$ of the implosion acceleration $\alpha$ is estimated by $\delta\alpha/\alpha \simeq \delta R/r_0 = (\delta R/R)(R/r_0) = \eta^{-1/3}(\delta R/R)$, where $r_0$ is the fuel initial radius and $\eta$ the density compression ratio. Typically the density compression ratio $\eta$ is about 1000 in inertial fusion. Then we obtain the relation of $\delta\alpha/\alpha \simeq \eta^{-1/3}\left[\{(\rho R)_0/\rho R\}^{1/2} - 1\right]$. In an inertial fusion reactor the degradation threshold of $(\rho R)/(\rho R)_0$ would be about 0.5 ~ 0.8, and $\delta\alpha/\alpha$ should be less than about 4.0% (Kawata et al., 1984). Based on this consideration, the driver beam illumination non-uniformity should be also mitigated to release a sufficient fusion energy output.

## 4. HOT-SPOT DYNAMICS

In order to investigate the fuel dynamics near the ignition, a hot-spot dynamics analysis is presented at the final stage of the fuel stagnation in this section. At the initial stage of the fuel target implosion the implosion non-uniformity reflects the imposed non-uniformity by, for example, the driver beam illumination non-uniformity, the target fabrication error and so on. During the implosion, the target radius shrinks till the fuel ignition, and the initial amplitude of the imposed non-uniformity grows as the fuel radius shrinks as discussed in Section 3. However, the previous implosion simulation results (Someya et al., 2006b) presented that the non-uniformity grows significantly and the non-uniformity amplitude enhancement is extraordinary especially around the ignition or final stagnation phase.

When the fuel behavior near the ignition is treated by a fluid model in the Lagrangian form, the Kelvin's theorem shows conservation of the vorticity $\omega$ (Landau et al., 1959):

$$\omega S = constant \qquad (1)$$

Here $S$ is the circulating area. In inertial fusion, the DT fuel mixing is induced by the R-T instability and also by the non-uniform implosion. The fuel mixing is one kind of the vortex, that is, the circulating motion. During the shrinkage of the DT fuel at the stagnation final stage just before the fuel ignition, $S$ would be reduced together with the shrinkage of the fuel radius scale length $L_{DT}$. Therefore, the vorciticy $\omega$ would be enhanced significantly at the final stage of the stagnation as follows:

$$\omega \sim \omega_0 (L_{DT0}/L_{DT})^2 \qquad (1)$$

In addition, the fusion fuel target shrinks in a 3 dimensional way. So in the Lagrangian frame the fuel mass is conserved during the stagnation, and the Ertel's thorem (Landau et al., 1959, Ertel, 1942) shows

$$\omega/\rho = \omega_0/\rho_0. \qquad (2)$$

In the inertial fusion fuel is compressed in 3D, and so

$$\omega \sim \omega_0 (L_{DT0}/L_{DT})^3. \qquad (3)$$

Based on this consideration, the vorticity $\omega$, that is, the circulation of the DT fuel would be enhanced significantly. The circulation enhancement induces the mixing of the cold fuel and hot fuel.

Due to the fuel mixing enhancement at the fuel stagnation phase, the fuel non-uniformity would be significantly enhanced at the final stage of the fuel compression just before the fuel ignition. In order to release the DT fusion energy stably, the initial non-uniformity, which is the seed of the consequent fuel mixing at the final stagnation phase.

## 5. SMOOTHING MECHANISMS OF BEAM ILLUMINATION NONUNIFORMITY

In inertial confinement fusion, the driver beam illumination non-uniformity leads a degradation of fusion energy output. Therefore, it is important to reduce the illumination non-uniformity. In this section two smoothing mechanisms are presented for the beam illumination non-uniformity in HIF.

As discussed above, HIB ions deposit their energy in a deep layer of the energy absorber, and so a density valley appears inside the energy absorber. Even in a direct driven fuel target shown in Fig. 1, a part of the HIB energy is converted to a radiation energy confined in the density valley. The density valley plays a role to confine the radiation energy and to smooth the HIB deposition energy partly. First we discuss the direct-indirect hybrid mode of the target implosion in HIF.

Then we introduce another smoothing mechanism by wobbling HIBs. In general a perturbation of physical quantity would be an origin of instability. Normally the perturbation phase is unknown so that the instability growth is discussed with the growth rate. However, if the perturbation phase is known, the instability growth can be controlled by a superposition of perturbations. If the perturbation is induced by, for example, a particle beam axis oscillation or wobbling, the perturbation phase could be controlled and the instability growth is mitigated by the superposition of the growing perturbations.

## 5.1 Direct-Indirect Hybrid Mode of Target Implosion

In order to realize an effective implosion, beam illumination non-uniformity on a fuel target must be suppressed less than a few percent. In this subsection a direct-indirect mixture implosion mode is discussed in heavy ion beam (HIB) inertial confinement fusion (HIF) in order to release sufficient fusion energy in a robust manner. In the direct-indirect mixture mode target, a low-density foam layer is inserted, and the radiation energy confinement is enhanced by the foam layer. In the foam layer the radiation transport is expected to smooth the HIB illumination non-uniformity in the lateral direction. Two-dimensional implosion simulations are performed (Someya et al., 2006a, Someya et al., 2006b), and show that the HIB illumination non-uniformity is well smoothed in the direct-indirect mixture target.

Figure 1 shows a typical fuel target in HIF. The radiation energy confined may smooth the HIB illumination non-uniformity. Therefore, we employ a foam layer to increase the confined radiation energy at the low density region as shown in Fig. 6. We call this target as a direct-indirect hybrid target. The mass density of the foam layer is 0.01 times the Al solid density in this study.

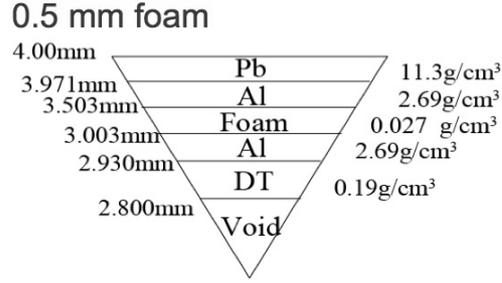

Fig. 4. Target structure with the 0.5 mm-thick foam. The foam inserted confines the radiation energy more to smooth the HIBs illumination nonuniformity.

The HIB pulse consists of a foot pulse and a main pulse as shown in Fig. 2. In this case, the total HIB energy is 4.0MJ. We employ a 32-HIBs illumination system (Skupsky et al., 1987). We evaluate the beam illumination non-uniformity at the target. In HF the Bragg peak deposition area plays the most important role for a target implosion. Therefore, we employ the total relative root-mean-square (RMS) as fallows;

$$\sigma_{RMS} = \sum_{i}^{n_r} w_i \sigma_{RMSi}, \quad \sigma_{RMSi} = \frac{1}{\langle E \rangle_i} \sqrt{\frac{\sum_{j}\sum_{k}(\langle E \rangle_i - E_{ijk})^2}{n_\theta n_\varphi}} \quad (6)$$

$$w_i = \frac{E_i}{E}$$

Here, $\sigma_{RMS}$ is the RMS non-uniformity of beam illumination. $\sigma_{RMSi}$ is the RMS non-uniformity on the $i$-th (r=constant) surface of deposition. $w_i$ is the weight function in order to include the Bragg peak effect. $n_r, n_\theta$ and $n_\phi$ are mesh numbers in each direction of the spherical coordinate. $\langle E \rangle_i$ is the mean deposition energy and $E_i$ is the total deposition energy on the *i*-th surface. $E$ is the total deposition energy. In this paper, two-dimensional (*r-θ*) simulations are performed, and the two-dimensional HIB-illumination time-dependent pattern at φ=90 deg is employed from the HIB illumination code.

In the foam layer the radiation transport is expected to smooth the HIB illumination non-uniformity in the lateral direction. To see the radiation transport effect on the implosion non-uniformity smoothing, we compare the results for the cases with the radiation transport (ON) and without the radiation transport (OFF) for the target shown in Fig. 6. Figure 7 presents the time histories of the RMS non-uniformity of the radiation temperature at the ablation front in the cases of the radiation transport ON and OFF. In Fig. 7 we see that the implosion non-uniformity at the ablation front becomes small effectively by the main pulse in the case of the radiation transport ON. During the main pulse, the implosion non-uniformity is well smoothed by the radiation transport effect.

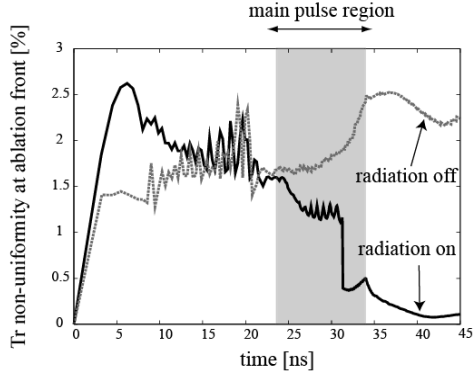

Fig. 5 The time histories of the RMS non-uniformity of the radiation temperature at the ablation front in the cases of the radiation transport ON and OFF.

In the direct-indirect hybrid implosion we employ the target with the 0.5mm-thickness foam as shown in Fig. 6. The peak conversion efficiencies of the HIB total energy to the radiation energy are ~ 4.5 % in the case of the 0.5mm foam and ~ 1.5 % in the case without the foam. The result means that a few hundreds kJ of the radiation energy is confined in the density valley and contributes to the nonuniformity mitigation. From these results, we find that the implosion mode in the case with the foam is a direct and indirect hybrid mode.

## 5.2 Wobbling Heavy Ion Beam Illumination

So far the dynamic stabilization for the R-T instability (Wolf, 1970, Troyon et al., 1971, Boris, 1977, Betti et al., 1993, Piritz et al., 2010, Piritz et al., 2011) has been discussed in order to obtain a uniform compression (Nuckolls et al., 1972, Atzeni et al., 2004) of a fusion fuel in inertial confinement fusion. The R-T dynamic stabilization was found many years ago (Wolf, 1970, Troyon et al., 1971) and is important in inertial fusion. It was found that the oscillation amplitude of the driving acceleration should be sufficiently large to stabilize the R-T instability (Wolf, 1970, Troyon et al., 1971, Boris, 1977, Betti et al., 1993, Piritz et al., 2010, Piritz et al., 2011). In inertial fusion, the fusion fuel compression is essentially important to reduce an input driver energy (Nuckolls et al., 1972, Atzeni et al., 2004), and the implosion uniformity is one of critical issues to compress the fusion fuel stably. Therefore, the R-T instability stabilization (Wolf, 1970, Troyon et al., 1971, Boris, 1977, Betti et al., 1993, Piritz et al., 2010, Piritz et al., 2011) or mitigation (Kawata et al., 1993, Kawata, 2012) is attractive to minimize the fusion fuel mix.

On the other hand, instabilities grow from a perturbation in general, and normally the perturbation phase is unknown. Therefore, we cannot control the perturbation phase, and usually the instability growth rate is discussed. However, if the perturbation phase is controlled and known, we can find a way to control the instability growth. One of the most typical and well-known mechanisms is the feedback control in which the perturbation phase is detected and the perturbation growth is controlled or mitigated or stabilized. In plasmas it is difficult to detect the perturbation phase and amplitude. However, even in plasmas, if we can actively impose the perturbation phase by the driving energy source wobbling or so, and therefore, if we know the phase of the perturbations, the perturbation growth can be controlled in a similar way (Kawata et al., 1993, Kawata, 2012). This control mechanism is apparently different from the dynamic stabilization shown in the previous works (Wolf, 1970, Troyon et al., 1971, Boris, 1977, Betti et al., 1993, Piritz et al., 2010, Piritz et al., 2011). For example, the growth of the Weibel instability or the filamentation instability (Weibel, 1959) driven by a particle beam or a jet could be controlled by the beam axis oscillation or wobbling. The oscillating and modulated beam induces the initial perturbation and also could define the perturbation phase. Therefore, the successive phase-defined perturbations are superposed, and we can use this property to mitigate the instability growth. Another example can be found in heavy ion beam inertial fusion; the heavy ion accelerator could have a capability to provide a beam axis wobbling with a high frequency. The wobbling heavy ion beams also define the perturbation phase. This means that the perturbation phase is known, and so successively imposed perturbations are superposed on plasma. We can use the capability to reduce the instability growth by the phase-controlled superposition of perturbations. In this subsection we discuss and clarify the dynamic mitigation mechanism for instabilities.

In heavy ion inertial fusion (HIF) the HIBs illumination non-uniformity would be mitigated by the wobbling beam motion, that is, the HIB axis oscillation or rotation (Kawata et al., 1993, Kawata, 2012).

In instabilities, one mode of an initial perturbation, from which an instability grows, may have the form of $a = a_0 e^{ikx+\gamma t}$, where $a_0$ is the amplitude, $k = 2\pi/\lambda$ is the wave number, $\lambda$ the wave length and $\gamma$ the growth rate of the instability. At $t=0$ the perturbation is imposed. The initial perturbation may grow by an onset of instability. After $\Delta t$, if the feedback control works on the system, another perturbation, which has an inverse phase with the detected amplitude at $t=0$, is actively imposed, so that the actual perturbation amplitude is mitigated very well. This is an ideal example for the instability mitigation.

In plasmas the perturbation phase and amplitude cannot be measured dynamically. However, by using a wobbling beam or an oscillating beam or a rotating beam or so, the initial perturbation is actively imposed so that the initial perturbation phase and amplitude are defined actively. In this case, the amplitude and phase of the unstable perturbation cannot be detected but can be defined by the input driver beam wobbling at least in the linear phase. In plasma it is difficult to realize the perfect feedback control, but a part of its idea can be adopted to the instability mitigation in plasmas. In actual, heavy ion beam accelerator can provide a controlled wobbling or oscillating beam with a high frequency.

If the energy driver beam wobbles uniformly in time, the imposed perturbation for a physical quantity of $F$ at $t = \tau$ may be written as

$$F = \delta F e^{i\Omega\tau} e^{\gamma(t-\tau)+i\vec{k}\cdot\vec{x}}. \qquad (7)$$

Here $\delta F$ is the amplitude, $\Omega$ the wobbling or oscillation frequency, and $\Omega\tau$ the phase shift of superposed perturbations. At each time $t = \tau$, the wobbler provides a new perturbation with the controlled phase shifted and amplitude defined by the driving wobbler itself. After the superposition of the perturbations, the overall perturbation is described as

$$\int_0^t d\tau\, \delta F e^{i\Omega\tau} e^{\gamma(t-\tau)+i\vec{k}\cdot\vec{x}} \propto \frac{\gamma+i\Omega}{\gamma^2+\Omega^2} \delta F e^{\gamma t} e^{i\vec{k}\cdot\vec{x}}. \qquad (8)$$

At each time of $t = \tau$ the driving wobbler provides a new perturbation with the shifted phase. Then each perturbation grows by the factor of $e^{\gamma t}$. At $t > \tau$ the superposed overall perturbation growth is modified as shown above. When $\Omega \gg \gamma$, the perturbation amplitude is reduced by the factor of $\gamma/\Omega$, compared with the pure

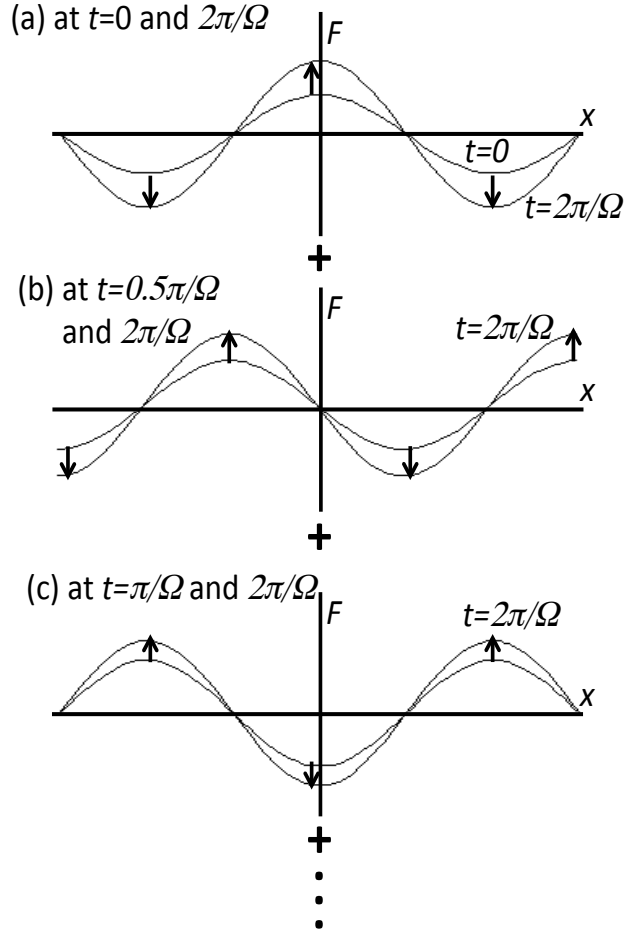

Fig. 6 Superposition of perturbations defined by the wobbling driver beam. At each time the wobbler provides a perturbation, whose amplitude and phase are defined by the wobbler itself. If the system is unstable, each perturbation is a source of instability. At a certain time the overall perturbation is the superposition of the growing perturbations. The superposed perturbation growth is mitigated by the beam wobbling motion.

instability growth ($\Omega = 0$) based on the energy deposition non-uniformity.

Figure 8 shows the perturbations decomposed, and each time the phase-defined perturbation is imposed actively by the driving wobbler. The perturbations are superposed at the time *t*. The wobbling trajectory is controlled by for example a beam accelerator or so, and the superposed perturbation phase and amplitude are controlled so that the overall perturbation growth is controlled.

From the analytical expression for the physical quantity *F* in Eq. (8), the mechanism proposed in this paper does not work, when $\Omega \ll \gamma$. Only for the modes, which satisfy the condition of $\Omega \geq \gamma$, the mechanism of the instability mitigation by the wobbler can be applied for its growth mitigation. For R-T instability, the growth rate $\gamma$ tends to become larger for a short wavelength. If $\Omega \ll \gamma$, the modes cannot be mitigated. In addition, if there are other sources of perturbations in the physical system and if the perturbation phase and amplitude are not controlled, this dynamic mitigation mechanism also does not work. For example, if the sphericity of an inertial fusion fuel target is degraded, the dynamic mitigation mechanism does not work. In this sense the dynamic mitigation mechanism is not almighty. Especially for a uniform compression of an inertial fusion fuel all the instability stabilization and mitigation mechanisms would contribute to release the fusion energy in stable.

Figure 9 shows an example simulation for R-T instability, which has one mode. In this example, two stratified fluids are superposed under an acceleration of $g = g_0 + \delta g$. In this example case the wobbling frequency $\Omega$ is $2\pi\gamma$, the amplitude of $\delta g$ is $0.1g_0$, and the results shown in Figs. 9 are at $t = 8\gamma$. In Fig. 9(a) $\delta g$ is constant and drives the R-T instability as usual, and in Fig. 9(b) the phase of $\delta g$ is shifted or oscillated with the frequency of $\Omega$ as stated above for the dynamic mitigation. The example simulation results also supports the effect of the dynamic mitigation mechanism well.

In HIF a fuel target is irradiated by HIBs, when the fuel target is injected and aligned at the center of the fusion reactor (Miyazawa et al., 2005, Someya et al., 2006b, Petzoldt, 1998). In this subsection, we employ (Pb$^+$) ion HIBs with the mean energy of 8GeV. The HIB temperature is 100MeV and the HIB transverse distribution is the Gaussian profile. The beam radius at the entrance of a fusion reactor is 35mm and the radius of a fusion reactor is 3m. We employ an Al monolayer pellet target structure with a 4.0mm external radius. The 32-HIBs positions are given as presented in (Skupsky, 1983). The HIBs illumination non-uniformity is evaluated by the global *rms* (see Eq. (1)), including also the Bragg peak effect in the energy deposition profile in the target radial direction. In this study, one HIB is divided into many beamlets, and the precise energy deposition

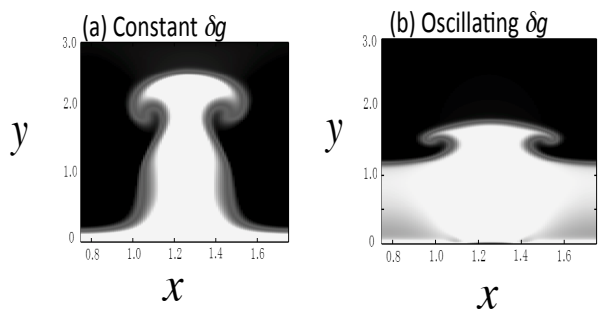

Fig. 7 Example simulation results for the R-T instability mitigation. (a) 10% acceleration nonuniformity drives the R-T instability as usual, and (b) 10% acceleration nonuniformity oscillates or wobbles.

is computed (Someya et al., 2004, Miyazawa et al., 2005, Ogoyski et al., 2010).

So far, we have found that the growth of the R-T instability would be mitigated well by a continuously vibrating non-uniformity acceleration field with a small amplitude compared with that of the averaged acceleration (Kawata et al., 1993, Kawata et al., 2009). It is realized by using a wobbling beam. Figure 10 shows a schematic diagram for the wobbling beam. However, in our previous work (Kawata et al., 2013) we found that at the initial stage of the wobbling HIBs illumination the illumination non-uniformity becomes huge and cannot be accepted for a stable fuel target implosion.

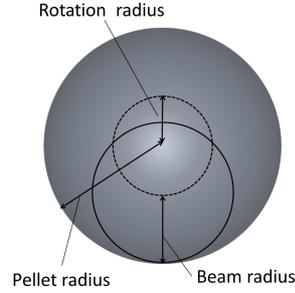

Fig. 8. Schematic diagram for a circularly wobbling beam

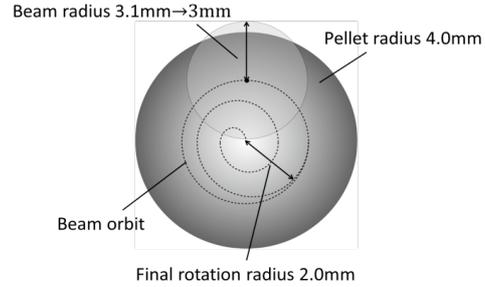

Fig. 9. Schematic diagram for spiral wobbling beam

This problem on the initial imprint of the rotating HIBs illumination is solved by the spiral wobbling HIBs as shown in Fig. 11. When the spirally wobbling beams in Fig. 11 are used, the initial imprint of the non-uniformity at the beginning of the irradiation is greatly reduced about from 14% to 4%. For the spiral wobbling beam the beam radius changes from 3.1mm to 3.0mm at $t = 1.3\tau_{wb}$. Here $\tau_{wb}$ is the time for one rotation of the wobbling beam axis. The beam rotation radius becomes 2.0mm at $t = 2.0\tau_{wb}$. After that, the beam rotation radius is 2.0mm. In this subsection, we employ the spirally wobbling beam for the HIBs illumination non-uniformity study.

Figure 12 shows the amplitude of the mode $(n, m) = (2, 0)$ vs. time, and Fig. 13 presents the spectrum of the mode (2, 0) in its frequency space. Here $(n, m)$ are the polar and azimuthal mode numbers, and $S_n^m$ is the amplitude of the spectrum, respectively. If the deposition energy distributed is perfectly spherically symmetric, the amplitude of the spectrum is 1.0 in the mode $(n, m) = (0, 0)$ in our study. For this reason, the amplitude of the mode $(n, m) = (0, 0)$ becomes large. As a result, the amplitude of spectrum mode $(n, m) = (2, 0)$ is largest and the mode $(n, m) = (2, 0)$ is dominant throughout the HIBs illumination. In Fig. 12 the time is normalized by the wobbling beam axis rotation time $\tau_{wb}$. In Fig. 13 $f_{wb}$ shows the wobbling HIBs rotation frequency. The result in Fig. 13 demonstrates that the small non-uniformity of the HIBs energy deposition has the oscillation with the same frequency and the double frequency with the wobbling HIBs oscillation frequency of $f_{wb}$.

## 6. CONCLUSIONS

The imploding fuel uniformity is one of the critical issues in HIF. First we discussed on the requirement for the implosion uniformity. The uniformity requirement is stringent. It is confirmed that the driver beam illumination non-uniformity should be less than a few %. The fuel deformation behavior is clarified by the 3D fluid model at the final stage of the fuel compression, and the nonlinear strong mode conversion is taken place so that the fuel large deformation appears just before the ignition.

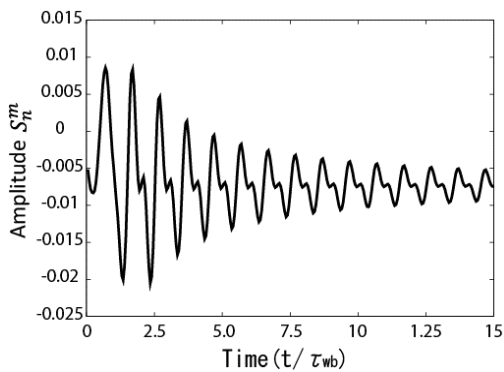

Fig. 10. The mode (2, 0) amplitude of HIBs nonuniformity vs. time

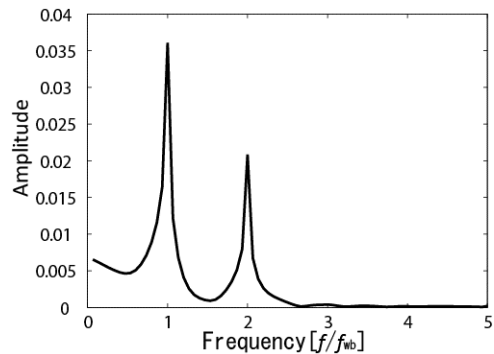

Fig. 11. Spectrum of the mode (2, 0) in its frequency space

The implosion dynamics demonstrated that the density gradient scale length is rather large, about several hundred μm. In order to reduce the HIB illumination non-uniformity, two non-uniformity mitigation mechanisms are also presented: the first one presented is the direct-indirect hybrid target, in which a part of the input HIBs energy is converted to the radiation energy in the density valley inside the target and the radiation energy contributes to the non-uniformity smoothing in HIF. The dynamic mitigation mechanism of instabilities is also introduced, and would contribute to smooth the beam driver non-uniformity. The HIB accelerator has a capability to control and rotate the HIB's axis precisely with a high frequency. The wobbling HIBs would induce the dynamic mitigation of the R-T instability in HIF. In this paper, we also found that the frequency spectrum of the HIBs illumination non-uniformity is synchronized with the rotation frequency of the wobbling beams. This result would work to reduce the growth of the R-T instability originated from the HIBs illumination non-uniformity. Finally we would like to stress that the dynamic mitigation mechanism shown in this paper is not almighty. It should be applied to the fusion fuel target compression together with other important mechanisms, for example, the dynamic stabilization, the density gradient mechanism, etc.

## ACKNOWLEDGEMENTS

The work was partly supported by MEXT (Ministry of Education, Culture, Sports, Science and Technology), JSPS (Japan Society for the Promotion of Science), CORE (Center for Optical Research and Education, Utsunomiya Univ., Japan), the program of High-End Foreign Expert in China, ASHULA project (ASian core program for High energy density science Using intense LAser photons), Japan-US fusion collaboration program, and ILE, Osaka University. The authors also appreciate Prof.

Yongtao Zhao and Prof. Dieter Hoffmann, the Chairmen of the 20[th] Int. Symposium on Heavy-Ion Inertial Fusion (HIF2014) for their perfect organization of HIF2014, Prof. K. Tanaka in Osaka University and Friends in U.S. Heavy Ion Fusion Virtual National Lab. for their fruitful discussions and supports.

**FIGURE CAPTIONS**

Figure 1. An example fuel target structure in heavy ion inertial fusion.

Figure 2. An input heavy ion beam pulse. The HIB pulse consists of the low power part (foot pulse) and the high power one (main pulse).

Figure 3. A mean density and a mean radiation temperature averaged over the θ direction at 36.2 nsec.

Figure 4. Target structure with the 0.5 mm-thick foam. The foam inserted confines the radiation energy more to smooth the HIBs illumination nonuniformity.

Figure 5. The time histories of the RMS non-uniformity of the radiation temperature at the ablation front in the cases of the radiation transport ON and OFF.

Figure 6. Superposition of perturbations defined by the wobbling driver beam. At each time the wobbler provides a perturbation, whose amplitude and phase are defined by the wobbler itself. If the system is unstable, each perturbation is a source of instability. At a certain time the overall perturbation is the superposition of the growing perturbations. The superposed perturbation growth is mitigated by the beam wobbling motion.

Figure 7. Example simulation results for the R-T instability mitigation. (a) 10% acceleration non-uniformity drives the R-T instability as usual, and (b) 10% acceleration non-uniformity oscillates or wobbles. The dynamic mitigation mechanism works well.

Figure 8. Schematic diagram for a circularly wobbling beam

Figure 9. Schematic diagram for spiral wobbling beam

Figure 10. The mode (2, 0) amplitude of HIBs non-uniformity vs. time

Figure 11. Spectrum of the mode (2, 0) in its frequency space